\documentclass[12pt]{article}
 
\usepackage {subeqn,cite}
 
\def\bib{\bibitem}
\def\be{\begin{equation}}
\def\ee{\end{equation}}
\def\barr{\begin{array}}
\def\earr{\end{array}}
\def\dis{\displaystyle}

\def\eg{ {\em e.g.}}

\def\etal{ {\em et al.}}

\def\lsim{\:\raisebox{-0.5ex}{$\stackrel{\textstyle<}{\sim}$}\:}
\def\gsim{\:\raisebox{-0.5ex}{$\stackrel{\textstyle>}{\sim}$}\:}
\def\mev{\: {\rm MeV} }
\def\gev{\: {\rm GeV} }

\def\fb{\: {\rm fb}}
\def\ra{\rightarrow}

\def\rp{$R_p \hspace{-1em}/\;\:$}

\def\ib#1,#2,#3{       {\em ibid.\/ }{\bf #1}, #3 (19#2)}
\def\ap#1,#2,#3{       {\em Ann.~Phys.~(NY)\/ }{\bf #1}, #3 (19#2)}
\def\appb#1,#2,#3{     {\em Acta Phys.\ Polon.\/ }{\bf B#1}, #3 (19#2)}
\def\cpc#1,#2,#3{      {\em Comput. Phys. Commun.\/ }{\bf #1}, #3 (19#2)}
\def\ijmp#1,#2,#3{     {\em Int.~J.~Mod.~Phys.\/ } {\bf A#1}, #3 (19#2)}
\def\mpl#1,#2,#3 {     {\em Mod.~Phys.~Lett.\/ } {\bf A#1}, #3 (19#2)}
\def\np#1,#2,#3{       {\em Nucl.~Phys.\/ }{\bf B#1}, #3 (19#2)}
\def\npps#1,#2,#3{     {\em Nucl.~Phys.~B (Proc.~Suppl.)\/ }
                             {\bf B#1}, #3 (19#2)}
\def\plb#1,#2,#3{      {\em Phys.~Lett.\/ }{\bf B#1}, #3 (19#2)}
\def\pr#1,#2,#3{       {\em Phys.~Rev.\/ }{\bf #1}, #3 (19#2)}
\def\prd#1,#2,#3{      {\em Phys.~Rev.\/ }{\bf D#1}, #3 (19#2)}
\def\prep#1,#2,#3{     {\em Phys.~Rep.\/ }{\bf #1}, #3 (19#2)}
\def\prl#1,#2,#3{      {\em Phys.~Rev.~Lett.\/ }{\bf #1}, #3 (19#2)}
\def\prog#1,#2,#3{     {\em Prog.~Theor.~Phys.\/ }{\bf #1}, #3 (19#2)}
\def\rmp#1,#2,#3{      {\em Rev.~Mod.~Phys.\/ }{\bf #1}, #3 (19#2)}
\def\sp#1,#2,#3{       {\em Sov.~Phys.-Usp.\/ }{\bf #1}, #3 (19#2)}
\def\zpc#1,#2,#3{      {\em Z. Phys.\/ }{\bf C#1}, #3 (19#2)}

\begin{document}

\setcounter{page}{0}
\thispagestyle{empty}
\renewcommand{\thefootnote}{\fnsymbol{footnote}}
\begin{flushright}
CERN-TH/97-51\\[2ex]
{\large \tt hep-ph/9703369} \\
\end{flushright}
\vskip 45pt
\begin{center}

\advance\baselineskip by 10pt
{\Large\bf Like-Sign Dileptons at the Fermilab Tevatron Revisited in 
the Light of the HERA High-{\boldmath $Q^2$} Anomaly} \\[2cm]

\advance\baselineskip by -10pt
{\bf Debajyoti Choudhury\footnote{debchou@mail.cern.ch}
        {\rm and}
    Sreerup Raychaudhuri\footnote{sreerup@mail.cern.ch}
}

\rm
\vspace{13pt}
{\em Theory Division, CERN, CH 1211 Geneva 23, Switzerland.}

\vspace{50pt}
{\bf Abstract}

\end{center}

\begin{quotation}
We re-examine like-sign dilepton signals at the Fermilab Tevatron assuming
that the excess high-$Q^2$ events recently seen at HERA are due to scalar
resonances such as squarks of $R$-parity-violating supersymmetry. For 
gluinos in the mass range of 200--350 GeV, the like-sign dilepton signal can 
help to make the crucial distinction between the most favoured squark 
explanation and other proposed solutions.
\end{quotation}

\vspace{10ex}

Pacs Nos.: 11.30.Pb, 14.80.Ly, 11.30.Fs, 14.80.-j \\

\vspace{10ex}
\noindent
CERN-TH/97-51\\
March 1997
\vfill
\newpage

\setcounter{footnote}{0}
\renewcommand{\thefootnote}{\arabic{footnote}}
\setcounter{page}{1}
\pagestyle{plain}
\advance \parskip by 10pt
\pagenumbering{arabic}

The particle physics community has recently been intrigued by reports of 
excess high-$Q^2$ events in both the H1~\cite{H1} and the ZEUS~\cite{ZEUS} 
detectors at HERA. 
Concentrated near a parton-momentum fraction of $x \sim 0.5$,
these events appear to point to some new 
effects~\cite{Adler,ChRa,DrMo,CERN_gang,DESY_gang,Blum,BKMW,HeRi,BCHZ}
beyond those expected in the Standard Model (SM).
Although the contrary opinion has also been expressed~\cite{Drees}, the 
jury is still out. Even assuming --- tentatively, in view of the low 
statistics --- that new physics has indeed been 
found, opinions differ as to what these effects could be. Various 
suggestions have been made, including compositeness~\cite{Adler}, 
contact interactions~\cite{BCHZ,CERN_gang}, 
and leptoquark~\cite{CERN_gang, DESY_gang, Blum, BKMW, HeRi} 
or squark~\cite{ChRa, DrMo, CERN_gang, DESY_gang}
resonances around 200 GeV. Of these, the last ---
namely squark resonances --- constitutes one of the more exciting 
possibilities, since these events could then be the first experimental 
signals of supersymmetry (SUSY) --- a desirable option for various 
theoretical reasons~\cite{mssm}. 

The possibility of a squark resonance at HERA was first considered by 
Hewett~\cite{Hewett} and subsequently by others~\cite{Kon}. This obviously 
requires a violation of $R$-parity~\cite{rpardef} in the form of 
lepton-number-violating $\lambda'$ operators~\cite{rpar}. The relevant 
Lagrangian is given by
\be
\barr{rcl}
{\cal L}_{\lambda'} & = & \dis
- \lambda'_{ijk}
      \left[  \tilde \nu_{iL} \overline{d_{kR}} d_{jL}
            + \tilde d_{jL} \overline{d_{kR}} \nu_{iL}
            + {\tilde d}_{kR}^\star \overline{(\nu_{iL})^c} d_{jL}
           \right.  \\[1.5ex]
      & & \dis \hspace*{3.2em}
      \left.
            - \tilde e_{iL} \overline{d_{kR}} u_{jL}
            - \tilde u_{jL} \overline{d_{kR}} e_{iL}
            - {\tilde d}^\star_{kR} \overline{(e_{iL})^c} u_{jL}
      \right]
       + {\rm h.c.} \ ,
\earr
          \label{lagr}
\ee
where the effects of quark mixing have been neglected. 
In a previous work~\cite{ChRa}, the present authors pointed out 
that, if the excess events at HERA are caused by a single 
dominant $\lambda'$ coupling, then the relevant ones are 
$\lambda'_{121}$ (with production of a 
$\tilde c_L$ squark), or $\lambda'_{131}$ (with production of a $\tilde t_L$ 
squark). A third possibility has also been suggested~\cite{CERN_gang, DrMo}
--- involving 
$\lambda'_{132}$, with a $\tilde t_L$ resonance. A value of $\lambda'_{1i1}$ 
in the interval 0.03--0.26 could give rise to the appropriate signal for the 
first two options. 
The third would require $\lambda'_{132}$ close to its experimental
upper bound of 0.3. Some other operators can contribute, though marginally,
for the highest experimentally allowed values of the relevant 
coupling constant~\cite{DrMo}.

A crucial feature of the $R$-parity-violating (\rp) signal is the fact, 
emphasized by all workers in this field, 
that the squark can also decay through 
$R$-parity-conserving channels to charginos and 
neutralinos, which could give rise to 
distinctive and unambiguous signals at HERA~\cite{H1RpV}. 
Observation of such a signal will be an immediate confirmation of the 
SUSY hypothesis. We understand that an intensive study of this nature 
is indeed under way~\cite{Sirois}. 
However, non-observation of such signals cannot rule out supersymmetry 
--- it will merely tell us that the direct $R$-parity-violating 
decay of the squark to $e^+ + d$ has to be the dominant
decay channel, while gaugino decay channels 
are suppressed by large gaugino masses or
 small couplings. In this case, of course, 
the decays of the squark will be identical with those of a leptoquark
with the same quantum numbers.

An interesting feature of the $R$-parity-violating SUSY solution is that 
low values of the $\lambda'$ coupling in question 
require large branching ratios to
the $R$-parity-violating decay channel $\tilde u_{iL} \rightarrow e^+ + d$.
While this may seem paradoxical at first, we need to recall that the 
number of excess events seen at HERA essentially scales as $\lambda'^2 
\beta$, where $\beta$ is the branching ratio of $\tilde u_{iL}$ to $e^+ + d$. 
Thus, small values of $\lambda'$ require a large branching ratio and vice versa. 
Such a scenario is far from unnatural --- it
can be achieved without difficulty by requiring 
the charginos and neutralinos to be either heavy or Higgsino-like, 
which merely corresponds to a restricted range
in the parameter space of the minimal supersymmetric Standard Model (MSSM).

In view of the above, it is now amusing to ask the following question. If, 
with more statistics from the 1997 run, the high-$Q^2$ anomaly does  
survive as a genuine effect and, moreover, as one compatible 
with a particle resonance, {\em but no further signals of SUSY are seen at 
HERA}, then the resources of that experiment will have been exhausted 
insofar as telling the nature of the new physics is concerned. It could be
a leptoquark or a squark with a small ($\lsim 0.1$) 
coupling. In this case, can any other running (or projected) experiment 
distinguish the supersymmetric option from the others?

The only other running high-energy facility with sufficient energy to 
produce the `particle' under consideration is the Fermilab Tevatron, running 
at a centre-of-mass energy of 1.8 TeV. Squarks or leptoquarks can
be pair-produced at this machine and their decays would give rise to 
distinct dilepton + multijet signals. For a squark mass of 200 GeV, 
we estimate the pair production cross section to be around 0.19 pb, which is 
consistent with Ref.~\cite{Spira}. Searches for such processes by the D0 
Collaboration have already put a bound of 175 GeV (with $\beta \simeq 1$) on 
the mass of the resonance at 95\% confidence level~\cite{D0_LQ}. 
This bound is based on a 
data sample of 117.7 pb$^{-1}$ and may be expected to increase to cover the 
mass range of interest ($\sim$ 180--220 GeV) as more data accumulates. If 
these searches fail to show up a resonance, it will be a sign of some serious 
glitch(es) in our understanding of the HERA anomaly, since most of the new 
physics solutions --- including SUSY --- do predict corresponding
signals at the Tevatron. On the other hand, if the dilepton + jets
signal is confirmed, it will inevitably lead us to the question of 
distinguishability between various theoretical models.

This report addresses the crucial question of distinguishability for the SUSY 
solution in the context 
of Fermilab Tevatron searches.  An interesting feature of SUSY is that the 
complete theory requires many more (light ?) particles than the squark. One 
of these is the gluino --- 
the fermionic superpartner of the gluon --- which is a 
strongly interacting {\em Majorana} fermion. As a result, relatively 
large cross sections from QCD production of  
gluinos are possible and processes involving them can violate fermion number,
leading to rather
 spectacular like-sign dilepton signatures. These have already 
been studied in some detail in the literature~\cite{BKT_95,Gu_Ro,Dr_Ro}
and our main purpose is to re-examine
their utility in the context of the squarks that can explain the HERA events.
The importance of this study lies in the fact that {\em such signals will be
present only if the HERA excess is caused by a squark resonance}.

We thus need to consider the pair-production of gluinos at the Fermilab 
Tevatron, followed by the decay of each gluino into a quark and a squark, 
where the squark is the 
same one that explains the HERA anomaly. Because of the Majorana nature 
of the gluinos, the squarks coming from their decay have equal probability 
of forming a like-sign or an unlike-sign pair. Subsequent decays of these 
squarks to $e^+ + d\:  (e^- + \bar d)$ would lead to 
equal numbers of like-sign and unlike-sign 
dilepton pairs. This decay chain differs from those studied in 
Refs.\cite{BKT_95,Gu_Ro} in that the squark now has no decays through 
gaugino-led channels, the latter being the exclusive decay modes 
studied therein.

Once produced, the gluino will decay into all possible quark--squark pairs 
that are kinematically allowed. 
Concentrating on the ($\lambda'_{121},\tilde c_L$)
scenario---the most promising one in the context of the HERA 
events~\cite{ChRa, DrMo, CERN_gang}---the decay channels of interest are 
\subequations
\be 
\tilde g \rightarrow \bar c + \tilde c_L \rightarrow \bar c + e^+ + d
\ee
and
\be 
\tilde g \rightarrow c + \tilde c_L^* \rightarrow c + e^- + \bar d \ ,
\ee
\endsubequations
which leads to a final state with a dilepton pair and jets.
The combinations $e^+ e^+$,  $e^+ e^-$, $e^- e^-$ appear with 
probability ratio 1:2:1. In other words, if $\tilde c_L$ were the 
only light squark, the decay of a gluino pair would lead to a like-sign 
dilepton final state in exactly half the cases. 

Now, the branching ratio of the gluino to $e^+ + X$ is given by 
\be
{\cal B} \equiv Br(\tilde g \ra e^+ + X) = \frac{ \sum_i 
          \beta_i ( 1 + x_i - \tilde x_i )  \lambda(1,x_i,\tilde x_i) }
              { \sum_i 
                 ( 1 + x_i - \tilde x_i )  \lambda(1,x_i,\tilde x_i) } \ ,
    \label{branching}
\ee
where $x_i \equiv m^2_{q_i} / M^2_{\tilde g}$, 
$\tilde x_i \equiv m^2_{\tilde q_i} / M^2_{\tilde g}$ 
and the sum over $i$ encompasses all squarks for which the decay
$\tilde g \rightarrow q_i \tilde q_i$ is kinematically possible. 
In Eq.(\ref{branching}), $\beta_i$ is 
the branching fraction of the relevant squark to $e^+ + X$, 
while $\lambda(a,b,c) \equiv \sqrt{(a - b - c)^2 - 4 b c}$. From our 
previous discussion, we see that, for $\lambda \lsim 0.1$, 
\[
  \beta_{\tilde c_L} \simeq 1, \quad  \beta_{\tilde s_L} \simeq 0, 
                               \quad \beta_{\tilde d_R} \simeq 0.5\ .
\]
As long as $\lambda'_{121}$ is the only non-zero \rp\ coupling, the 
cascade deacys of the other squarks will result in final states with 
$e^\pm$ or (anti-)neutrinos. Such 
event topologies have been considered in Refs.\cite{BKT_95,Gu_Ro}. 
Rather than repeat their analyses, which are quite comprehensive for
these decay chains, we shall make the {\em conservative}
assumption that $\beta_i = 0$ for these squarks.
With this simplification, ${\cal B}$ in Eq.(\ref{branching}) is determined 
solely by the gluino and the squark masses. We find that for 
$\lambda'_{121} \simeq 0.1$, $ {\cal B} $ is 0.54 or less.

As the masses of the squarks are unknown parameters in the MSSM, we have to 
make some assumptions for the sake of simplicity. An explanation of the 
HERA anomaly requires only two squarks to be light. These are the 
$\tilde c_L$ (obviously) and
the $\tilde s_L$ whose mass is given by the relation
\be
        m^2_{\tilde s_L} = m^2_{\tilde c_L} - m_c^2 + m_s^2 
                         - m_W^2 \cos 2 \beta    \ .
         \label{sq_mass_split}
\ee
Numerically, for $m_{\tilde c_L} \simeq 200$ GeV, 
we find that $m_{\tilde s_L}$ varies in the range 200--215 GeV 
as $\tan \beta$ varies from 1 to 50. The other 
squarks can be as heavy (or as light) as we please. We make the ad hoc
assumption that 
\be
     m_{\tilde u_{Ri} } = m_{\tilde d_{Ri} } = m_{\tilde u_{L \alpha} } 
                        \equiv m_{\tilde q }  \ ,
\ee
where $i = 1,2,3$ and $\alpha = 1,3$. 
The masses of $\tilde d_L$ and $\tilde b_L$
are given by relations analogous to that in Eq.(\ref{sq_mass_split}). 
This is not perhaps the optimal spectrum for the signal in question, 
nor is it the one most favoured by model-builders, but it is 
certainly the simplest one. We have checked that the qualitative 
features of our results do not change with different
Ans\"atze for the squark masses, although the detailed numerics are somewhat 
affected~\cite{dR_expl}.

\begin{figure}[h]
\vskip 3.9in
      \relax\noindent\hskip -1.0in\relax{\includegraphics{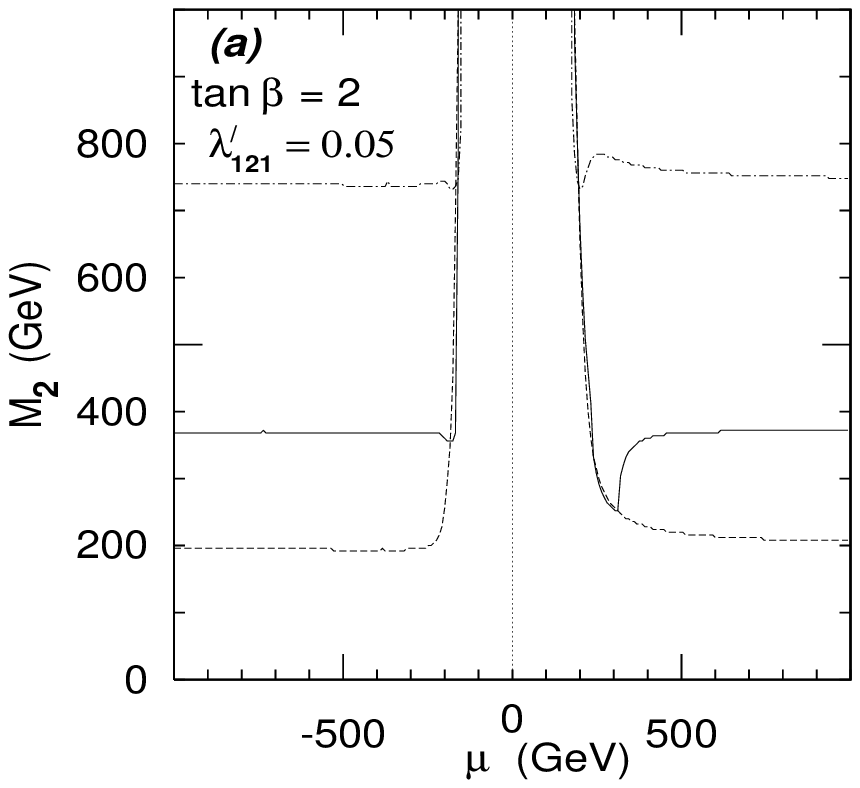}}
      \hskip 2.8in\relax{\includegraphics{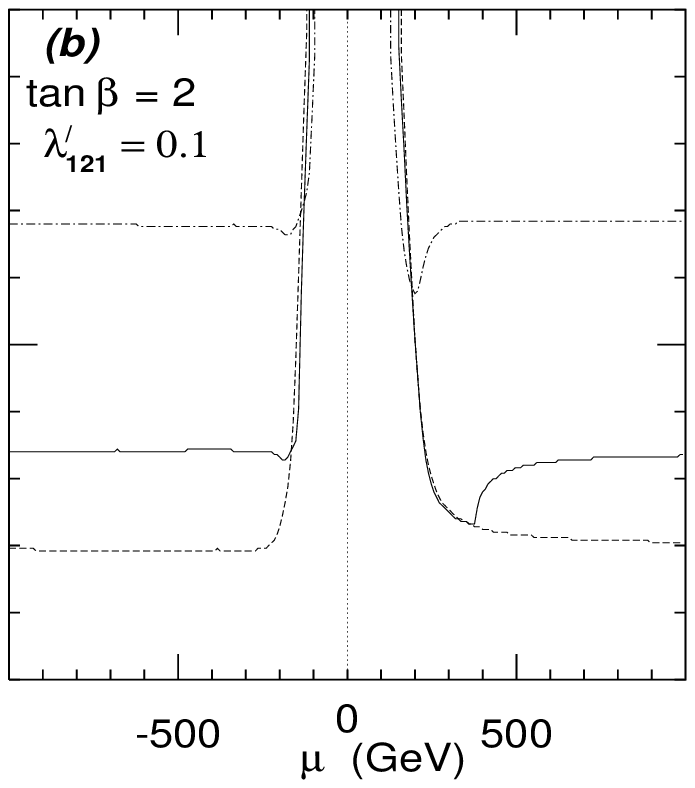}} 
\vspace{-11.5ex}
\caption{\em Contours for $\beta_{\tilde c_L} = 0.9$ in the $M_2$--$\mu$ 
             plane for $\tan \beta = 2$ and various values of the ratio 
             $M_1/ M_2$. Dashed, solid and dot-dashed lines correspond to 
             $M_1/M_2 = 2, 0.5, 0.25$ respectively. The dependence on 
             $\tan \beta$ is not very significant.}
              \label{fig:branching}
\end{figure}
Since we want the \rp\ channel to dominate the squark decay width, the gluino
must be heavier than the squark. And so must be the charginos and neutralinos, 
unless of course the lighter ones are Higgsino-dominated. Within the framework 
of gaugino mass unification, and the LEP constraints, this 
implies a large value for the gaugino mass parameter $M_2$. Within the same 
framework, this would further
imply very large gluino masses (generally $\gsim 500 \gev$) 
and hence extremely tiny 
rates at the Tevatron. Note, however, that gaugino mass unification is not 
very well explored within the context of $R$-parity violation. While Grand 
Unification can perhaps be made consistent with \rp~\cite{RP_GUTS}, such 
models typically need extra multiplets, all consequences of 
which are not very well studied. Hence, we shall {\em not} assume gaugino-mass 
unification. The squark mass spectrum does not assume scalar mass unification
either. We thus adopt a purely phenomenological approach to the SUSY mass 
spectrum.

Once we give up gaugino mass unification as a hypothesis, the soft
supersymmetry-breaking parameters $M_1,M_2,M_3$ of the MSSM are no longer 
related. It is 
then possible to obtain large branching ratios for $\tilde c_L \rightarrow
e^+ + d$ for a fairly large range of the parameter space. 
This is illustrated in the $M_2$--$\mu$ plane for a fixed value 
$\tan \beta = 2$ in Fig.{fig:branching}. We have chosen two values of the
coupling (a) $\lambda'_{121} = 0.05$ and 
(b) $\lambda'_{121} = 0.1$, both of which require a large branching 
ratio in order to explain the HERA events (see Fig. 2 of Ref.\cite{ChRa}). 
The region above each contour corresponds to  
$\beta_{\tilde c_L} \geq 0.9$. Dashed, solid and dot-dashed lines 
correspond to the ratios $M_1/M_2 = 2, 0.5, 0.25$ respectively, of 
which the second approximates closely to the ratio $M_1 /M_2 = 
\frac{5}{3} (g_1/g_2)^2$ obtained by assuming gaugino mass 
unification purely in the electroweak sector of the MSSM.
It is also worth mentioning that for this figure we have
included LEP constraints on the \rp\ scenario in question --- these
come only from the total width of the $Z$, to which the 
excess SUSY contribution is ~\cite{LEPEWG}:
\be
      \sum_{i,j=1}^2 \Gamma( Z \ra \tilde \chi_i^+ \tilde \chi_j^-)
          + \sum_{i,j=1}^4 \Gamma( Z \ra \tilde \chi_i^0 \tilde \chi_j^0)
               \lsim 23.1 \mev \ .
          \label{lep_constr}
\ee
It is at once apparent that somewhat large values of $M_2$ (and $M_1$)
are demanded. Low values of $|\mu |$ are disallowed,
partly by the LEP constraint and partly because they correspond to
small branching ratios. Finally we should note that varying 
$\tan \beta$ does not change the contours much. Accordingly, we have
fixed $\tan \beta = 2$.  Our plots are in good agreement with those of 
Ref.\cite{CERN_gang}. 
 
The like-sign dilepton signal has been estimated assuming   
$\beta_{\tilde c_L} = 1$ and using a parton-level Monte Carlo event 
generator. Our estimates for gluino pair-production agree well with 
those of Ref.\cite{Spira}.
For our simulation, we use the CTEQ3M structure functions~\cite{CTEQ},
which were calculated using the package PDFLIB~\cite{PDFLIB}. 
We assume a detector rapidity coverage of 3 for both jets 
and leptons : 
\be
     | \eta(\ell) |, | \eta(j) |  < 3 \ .
               \label{rap_cov}
\ee
All putative jets that lie within a cone of 
$\Delta R \equiv \sqrt{ \Delta \eta^2 + \Delta \phi^2} \leq 0.7$ of each other 
are merged (the momenta added vectorially) to form a single jet. Here 
$\Delta \eta$ is the difference of their rapidities and $\Delta \phi$ denotes 
their azimuthal separation. 
The major SM backgrounds to the like-sign dilepton signal emanate from 
$t \bar t$ and $W Z$ production~\cite{BKT_95, Gu_Ro}. These can be severely 
suppressed by using suitable kinematical cuts. We demand that the leptons
have sufficient transverse momentum and be relatively isolated :
\be
            p_T(\ell) > 15 \gev, \qquad E_T^{\rm ac}(\ell) < p_T(\ell) / 4 \ ,
     \label{pT_iso_cut}
\ee
where $ E_T^{\rm ac}(\ell)$ denotes the total hadronic energy within a cone 
of $\Delta R = 0.4$ around the lepton $\ell$~\cite{Cut_comment}. With the
imposition of these cuts, the SM background due to $t \bar t $ production
was estimated to be as low as $0.3 \fb$, while 
$2.1 \fb$ come from $WZ$ production~\cite{BKT_95}. 
Clearly, the background level 
is very low and thus the discovery limit is primarily governed by the 
signal strength. Our results are consistent with those of Ref.~\cite{Gu_Ro},
which were checked against an ISAJET-based calculation with reliable cuts and 
efficiencies.

\begin{figure}[htb]
\vskip 3.9in
      \relax\noindent\hskip 0.5in\relax{\includegraphics{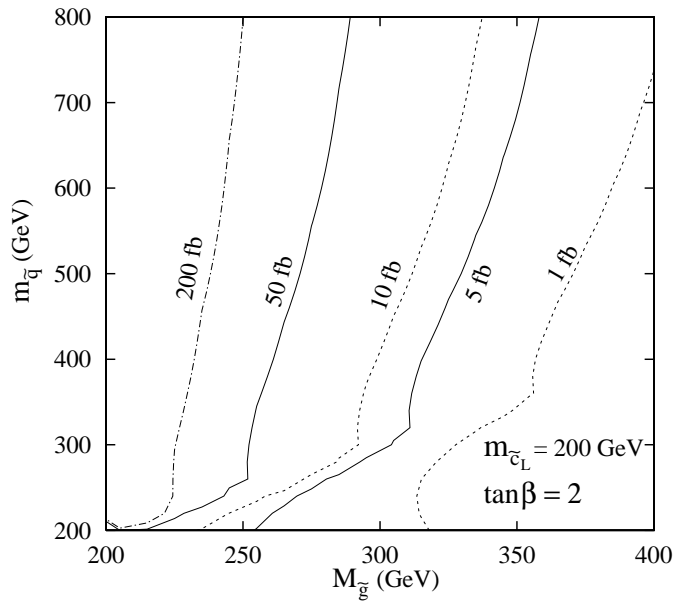}}
\vspace{-11.5ex}
\caption{\em Contours for the like-sign dilepton cross-section at the 
             Tevatron in the $m_{\tilde g}$--$m_{\tilde q}$ plane. The 
             dependence on $\tan \beta$ is minimal. Changes in $m_{\tilde c_L}$
             affect only the lower edge of the curve (see text).}
              \label{fig:simul}
\end{figure}
In Fig.~\ref{fig:simul}, we present contours of cross-section for the like-sign
dilepton signal in the plane of $m_{\tilde q}$, the common squark mass, and 
$M_{\tilde g}$, the gluino mass. We have fixed the mass of the $\tilde c_L$ at 
200 GeV and $\tan \beta = 2$ 
(which fixes the mass of $\tilde s_L \simeq 210 \gev$). 
We have checked that a variation of the 
former between 180 and 220 GeV and the latter
between 1 and 50 makes little visual impact on the figure, 
although the detailed
numbers do change slightly. It should be noted, though, that 
the lower edge of the curves represent {\em conservative} estimates 
as we have neglected the cascade decays of the squarks into the dilepton 
channel. Inclusion of these is expected to make the low-$m_{\tilde q}$ region
more accessible~\cite{BKT_95,Gu_Ro}. 
Dotted lines in the figure indicate the 
kinematic reaches of the Tevatron with 0.1 fb$^{-1}$ and 1 fb$^{-1}$ of 
integrated luminosity. 
In view of the low background of $\sim 2.4$ fb, we show contours of 5 fb and
50 fb, since these imply 5 events at the Tevatron with, respectively,
0.1 fb$^{-1}$ of data (which are currently available) 
and with 1 fb$^{-1}$ of data (which should 
become available with the Main Injector run). 
The solid contours may, therefore, be taken as discovery limits. 
The dot-dashed line corresponds to a signal size of 200 fb, which is perhaps
rather optimistic. It is immediately obvious that if Nature favours us 
with a somewhat low-lying gluino state ($< 350$ GeV), then it 
should be possible to observe 
like-sign dilepton signals at the Tevatron, unless, 
indeed, there also happens to be a bunch of low-lying squark states around 
200 GeV apart from the ($\tilde c_L, \tilde s_L$)
pair which have $\beta_i = 0$ (thereby reducing ${\cal B}$). 
On the other hand, if these signals are not seen, it should be 
possible to rule out a substantial 
region in the squark--gluino mass plane, which 
has not been hitherto accessible to Tevatron searches (assuming, of course, 
that the HERA events are susceptible to a SUSY explanation with a
$\tilde c_L$ resonance).

We now comment on the equally interesting possibility that the HERA events 
are due to 
a stop $\tilde t_L$ resonance through a $\lambda'_{131}$ (or $\lambda'_{121}$) 
coupling.
In this case, a strong like-sign dilepton signal 
requires the gluino to decay to a top--stop
pair which is only possible when the gluino mass is about 370 GeV or more.
This mass range is at the kinematic limit of the Tevatron and, as 
Fig.~\ref{fig:simul} makes clear, yields a very small signal.
Thus, we conclude that the top-stop decay mode of the gluino will not
yield an observable like-sign dilepton signal through the direct \rp\ decay 
mode. However, the story
does not end here, because the presence of an $\sim 200$ GeV $\tilde t_L$
implies the presence of a light $\tilde b_L$. Recalling that the masses
are related by 
\[
        m^2_{\tilde b_L} = m^2_{\tilde t_L} - m_t^2 + m_b^2
                         - m_W^2 \cos 2 \beta    \ ,
\]
it is easy to check that a 200 GeV $\tilde t_L$ implies a $\tilde b_L$
in the mass range 105--132 GeV. (The lower part of this range is ruled
out in the limit of vanishing left-right stop mixing by constraints from the 
$\rho$ parameter, although in the presence of mixing this constraint can be 
relaxed~\cite{ChRa}.) Unfortunately, 
even though the gluino can decay dominantly to 
the $\tilde b_L$, the latter (like the $\tilde s_L$)
does not decay to charged leptons and will not contribute to
the like-sign dilepton signal. On the other hand, if there is a low-lying
$\tilde d_R$, one could still get a signal, although even this is 
suppressed by the fact that $\beta_{\tilde d_R} \leq 0.5$. We do
not go into this aspect any further since there is no compelling reason 
at the moment to postulate a light $\tilde d_R$. Thus we are forced to the 
conclusion that the like-sign dilepton signal is unlikely 
to be able to distinguish the stop solution to the HERA anomaly
from the possible non-supersymmetric alternatives. However, a $\tilde b_L$
of mass upto 120--130 GeV is likely to be accessible to the third-generation
leptoquark search at the Tevatron~\cite{D0_LQ} (in the $b\bar b$ + missing
$E_T$ channnel) in the event of a luminosity upgrade. It might also be possible
to access a part of the parameter space through cascade 
decays~\cite{BKT_95,Gu_Ro}.

To summarize, then, we have investigated the possibility of observing
like-sign dileptons at the Fermilab Tevatron in view of the possibility
--- no longer so remote, in view of the HERA anomaly --- that there
exists a $\tilde c_L$ squark of mass around 200 GeV. If there is also 
a gluino in the 200--350 GeV range, it might be possible to see such
signals at the Tevatron, if not now, then in the near future, with the
projected luminosity upgrade. Even with the present data sample
it should be possible to explore a hitherto untouched part of the 
parameter space, at least for the scenario in question. Unfortunately,
the alternative SUSY solution of the HERA anomaly 
with a 200 GeV $\tilde t_L$
cannot be addressed well through this particular signal. Even so, much can
be achieved by searching for like-sign dileptons in the present data
sample and we would urge upon our experimental colleagues the importance of
carrying out such searches with the highest priority.

The authors would like to thank Manuel Drees and N.K. Mondal (D0 Collaboration)
for discussions.

\end{document}